# Digital Twin: From Concept to Practice

Ashwin Agrawal, M.S.[1]; Martin Fischer, Ph.D.[2]; Vishal Singh, Ph.D.[3]

[1]Department of Civil and Environmental Engineering, Stanford University, Stanford, CA, USA

(Corresponding author) Email: ashwin15@stanford.edu

[2]Professor, Civil and Environmental Engineering, Stanford University, Stanford, CA, USA

Email: fischer@stanford.edu

[3]Associate Professor, Centre of Product Design and Manufacturing, Indian Institute of Science, Bangalore, India, Email: singhv@iisc.ac.in

## ABSTRACT

Recent technological developments and advances in Artificial Intelligence (AI) have enabled sophisticated capabilities to be a part of Digital Twin (DT), virtually making it possible to introduce automation into all aspects of work processes. Given these possibilities that DT can offer, practitioners are facing increasingly difficult decisions regarding what capabilities to select while deploying a DT in practice. The lack of research in this field has not helped either. It has resulted in the rebranding and reuse of emerging technological capabilities like prediction, simulation, AI, and Machine Learning (ML) as necessary constituents of DT. Inappropriate selection of capabilities in a DT can result in missed opportunities, strategic misalignments, inflated expectations, and risk of it being rejected as just hype by the practitioners. To alleviate this challenge, this paper proposes the digitalization framework, designed and developed by following a Design Science Research (DSR) methodology over a period of 18 months. The framework can help practitioners select an appropriate level of sophistication in a DT by weighing the pros and cons for each level, deciding evaluation criteria for the digital twin system, and assessing the implications of the selected DT on the organizational processes and strategies, and value creation. Three real-life case studies illustrate the application and usefulness of the framework.

# 1. INTRODUCTION

Digitalization offers numerous possibilities to improve performance and productivity within the Architecture, Engineering, and Construction (AEC) industry (Hampson and Tatum 1993). One such technology that has gotten a lot of attention recently is Digital Twin (DT) (Boje et al. 2020; Grieves and Vickers 2017). It promises to give a multi-dimensional view of how an asset will perform by simulating, predicting, and informing decisions based on real-world conditions (Autodesk 2021). A recent Gartner survey reveals that by 2022, over two-thirds of the companies that have implemented sensor technology anticipate to have deployed at least one DT in production (Gartner 2019).

However, the problem is that there is no so-called *"universal DT"* that everyone can deploy. DT comes with a wide variety of sophisticated capabilities ranging from as simple as a digital representation (Canedo 2016; Schroeder et al. 2016) to increasingly complex models with predictive and prescriptive capabilities (Gabor et al. 2016; Glaessgen and Stargel 2012). Naturally, the technological capabilities, resources needed to build DT, and the value that DT adds to the business would differ in every case as well. Therefore, for a successful deployment of a DT, the managers and practitioners would need to select an appropriate level of sophistication in a DT, articulate the technological requirements to build it, and clearly communicate the strategic vision for its implementation to the top management.

But given these varied possibilities that DT can offer, practitioners themselves are confused, and face increasingly difficult decisions regarding what type of technological capabilities to select in a DT while deploying it in the AEC industry (Shao and Helu 2020; Feng et al. 2020; Agrawal et al. 2022). A lack of understanding in practitioners and the company's management regarding the type of capabilities needed in DT can result in unrealistic hopes from the technology (Love et al. 2020), strategic misalignments (Hampson and Tatum 1993), misallocation of resources, inability to realize benefits from the technology (Love and Matthews 2019), and ultimately a rejection of DT as hype (Wright and Davidson 2020).

This paper answers the following research question: *"Given the wide range of possibilities that DT can offer; how should practitioners select an appropriate level of sophistication in a DT to deploy in practice?"* Specifically, the paper aims to facilitate this process of selection for practitioners by proposing the digitalization framework. This framework highlights two perspectives that should be kept in mind while selecting the sophistication in a DT: (1) business value that the company expects from DT deployment, and (2) technological capabilities the company possesses to develop a DT. The framework further helps to align these two perspectives, thus helping practitioners evaluate and understand the different forces in play while deploying a DT in practice.

In addition to facilitating the selection of an appropriate level of sophistication in a DT, the digitalization framework helps managers and practitioners understand and highlight various strategic misalignments in the deployment of DT; inculcate a strategic mindset within the organization; and set up a long-term strategic vision/roadmap for digitalization in the company. Educators and researchers will arguably find value in the highlighted dichotomy between the business value that practitioners aspire from DT and the technological capabilities they possess to build it. Awareness of this dichotomy that practitioners regularly face in practice would enable the researchers develop methods and practices that are more likely to succeed in the actual field deployment of a DT. We thus aspire that our digitalization framework, to some extent, can accelerate and steer the adoption of digital technologies in the right direction, which still has been lagging in the AEC industry.

The paper starts by reviewing the literature in Section 2. Section 3 provides the research and validation method used to develop the framework. This is followed by introducing the digitalization framework in Section 4. In Section 5, three real-life case studies showcase the relevance of the digitalization framework. The paper concludes by discussing the findings and their implications for the AEC industry in Section 6.

## 2. LITERATURE REVIEW

This section first reviews the research context, DT, in sub-section 2.1 and levels of DT, in sub-section 2.2. It then focuses on the studies most relevant to this work's focus, methods to select the appropriate level of DT, in sub-section 2.3. Finally, the observed gaps in the literature are summarized in sub-section 2.4.

### 2.1 What is a DT?

The concept of a physical twin, a precursor to DT, is rather old and dates to NASA's Apollo program (Schleich et al. 2017). Identical space vehicles were built, and one vehicle remained on earth called the twin. The twin was used to mirror the precise in-flight conditions, run simulations, and thus assist the astronauts with the best possible solution. Therefore, the idea of a "twin" broadly covers all the prototypes that help to mirror the actual operating conditions.

Naturally, it is costly and almost impractical to construct a physical twin of every asset/entity. Therefore, the idea of physical twinning was extended further to construct a twin digitally. The proposition of "digitally" twinning an asset/entity that helps to mirror the actual operating conditions but at the same time is less costly and practical is precisely the motivation for the DT concept.

Michael Grieves first presented the idea of DT in 2003 as a digital information construct of the physical system, which optimally includes all the relevant information required to complete the task at hand and is linked with the physical system in question (Grieves and Vickers 2017). The main DT components (as shown in Figure 1) are the: (1) physical component, (2) virtual model, and (3) data that connects the components. The data flow from the physical component to the virtual model is raw and requires processing to convert to helpful information. On the other hand, the data flow from the virtual to the physical world is processed information that can be used to manage the day-to-day usage of the physical entity.

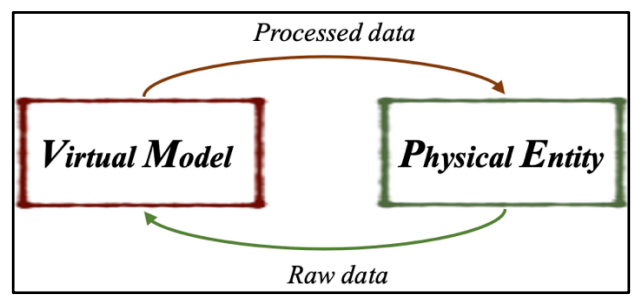

*Figure 1 Digital Twin Paradigm*

The first formal definition of DT was coined by NASA in 2012 as (Glaessgen and Stargel 2012): "an integrated multi-physics, multi-scale, probabilistic simulation of a vehicle or system that uses the best available physical models, sensor updates, fleet history, etc., to mirror the life of its flying twin." Although the concept originated from the aerospace industry, owing to its usefulness it has spread in many other industries like manufacturing and construction. The following paragraphs summarize various DT applications, implementation themes, and barriers to adaption in these industries.

In the manufacturing industry, the primary purpose of DT has been to represent the system's complex behavior, considering the possible consequences of external factors, human interactions, and design constraints (Rosen et al. 2015; Gabor et al. 2016). Kritzinger et al. (2018) reviewed over 40 articles on DT application in the manufacturing industry and categorized the focus areas of DT implementation in five specific categories apart from the general manufacturing applications: (1) layout planning for automated production planning and evaluation (Uhlemann et al. 2017), (2) optimization of the product lifecycle (Boschert and Rosen 2016), (3) production planning and control to improve and automate decision support (Rosen et al. 2015), (4) manufacturing process re-design (Schleich et al. 2017), and (5) predicting and managing maintenance (Susto et al. 2015). Cimino et al. (2019) also reviewed over 50 articles on DT applications in the manufacturing industry and found a similar categorization for DT focus areas in the manufacturing industry.

Based on reviews done by (Opoku et al. 2021; Al-Sehrawy and Kumar 2021; Jiang et al. 2021), applications of DT have been demonstrated throughout an asset lifecycle in the AEC industry. DT has been implemented in the design and engineering phase by using a combination of Building Information Modelling (BIM) and Wireless Sensor Network (WSN) to provide designers with efficient real-time information during project design (Lin and Cheung 2020). Du et al. (2020) introduces the concept of a cognition DT, which can provide selective and personalized information to designers and engineers, thereby reducing information overload and improving efficiency. Martinelli et al. (2019) used a DT for cost estimation during the preliminary conceptual design phase. In the construction phase, DT has been used for construction progress monitoring and management (Bueno et al. 2018), construction quality and safety monitoring (Akula et al. 2013), and machine and material monitoring (Zhou et al. 2019). In the Operations and Maintenance (O&M) phase, DT combined with ML has been used for simplified analysis of energy usage in buildings (Austin et al. 2020). Francisco et al. (2020) ideates a DT enabled urban energy management platforms by enabling identification of building retrofit strategies. DT has also shown to be effective in data querying and supporting decision-making during the O&M phase (Q. Lu et al. 2020). Several other applications of DT have also been demonstrated outside asset management space in the AEC industry, like enhanced risk-informed decision making (Ham and Kim 2020), and disaster management (Ford and Wolf 2020; Fan et al. 2020).

Although the above-described pilot applications of DT are motivating, several barriers need to be resolved to enable a widespread adaption of DT. Neto et al. (2020) finds from expert interviews that lack of structured project pathways to implement DT, and organization culture and strategy resistant to change are the major blockers for DT adaption, highlighting a lack of strategic vision in practitioners and organizations. Wache and Dinter (2020) reviews the DT literature and finds that the current literature solely focuses on technology deployment and underrepresents managerial/organizational point of view, critical for DT adaption. Perno et al. (2022) reviews over 40 articles on DT implementation and finds that difficulty in making suitable decisions and investments regarding the enabling technologies (Ezhilarasu et al. 2019) and difficulty in identifying clear value propositions associated with DT solutions (Wishnow et al. 2019) as the barriers for its adaption in the industry.

The digitalization framework, to some extent, can help alleviate the above problems. It helps the managers make suitable decisions regarding the enabling technologies by selecting an appropriate level of sophistication in a DT. Moreover, it also forces the practitioners to clearly articulate the value proposition of DT by inculcating a strategic mindset and highlighting various strategic misalignments. Finally, the digitalization framework helps create a strategic roadmap thus addressing the under-representation of managerial perspective. More details on this have been provided in Section 6.

## 2.2 Existing works on levels of DT

To select an appropriate sophistication in a DT, it is first essential to specify the different levels of sophistication in a DT. The levels of DT exactly provide the answer to this question and thus describe and compare the different types of DT. The digitalization framework builds upon the existing models for levels of DT and helps to select the appropriate level to be deployed in practice. Below we discuss some of the existing work around levels of DT. Although we showcase the use of the digitalization framework with the levels of DT hierarchy proposed by Gartner, our framework is broadly applicable to any of these hierarchies.

Three different types of DT have been described by (Kritzinger et al. 2018) as: digital model, digital shadow, and digital twin, depending on the amount of data integration. Madni et al. (2019) describes the levels of DT maturity as: pre-digital twin, digital twin, adaptive digital twin, and intelligent digital twin. Agrawal et al. (2021) provides a two-dimensional framework for levels of DT based on the intelligence capability at each level. Autodesk (2021) also describes a 5-level hierarchy as: (1) descriptive twin, a visual replica of the asset, (2) informative twin, captures and aggregates defined data, (3) predictive twin, use operational data to gain future insights, (4) comprehensive twin, generates 'what-if' scenarios, and (5) autonomous twin, acts on behalf of the users. Gartner (2013) describes different levels of digital analytics capabilities that can be present as: descriptive, diagnostic, predictive, and prescriptive abilities.

The taxonomy provided by Gartner is one of the most observed hierarchies in the literature. For example, a lot of similarities can be observed between the hierarchy proposed by Gartner and Autodesk. Davenport and Harris (2017) offers a very similar hierarchy consisting of statistical analysis to answer 'why' questions (like analysis level of Gartner), forecasting and predictive modeling (prediction level in Gartner), and optimization level (the equivalent of prescription level). Pyne et al. (2016) also define the level of data analytics into three levels: description, prediction, and prescription. Nguyen et al. (2018) reviews over 80 articles on the application of big data analytics in different domains like manufacturing, procurement, and logistics management to report that the prediction and prescription levels have been found in over 40% of the papers, again reinforcing the immense popularity of this taxonomy. Hence, for the scope of this paper, we use Gartner's taxonomy in the digitalization framework. Below we provide a summary of the taxonomy proposed by Gartner.

The rationale of using Gartner's taxonomy is to examine the extent to which DT is being used to support the decision-making processes (see Figure 2). At the lowest level, DT only represents the information in a useful form (description). Humans can use this information to make the final decision. At a higher level, DT goes a step further and reasons why something might be happening (diagnostic). Further, to evaluate different alternatives, at the highest levels, DT tries to predict the possible future outcome (prediction) and decide the action based on the objectives and preferences (prescription).

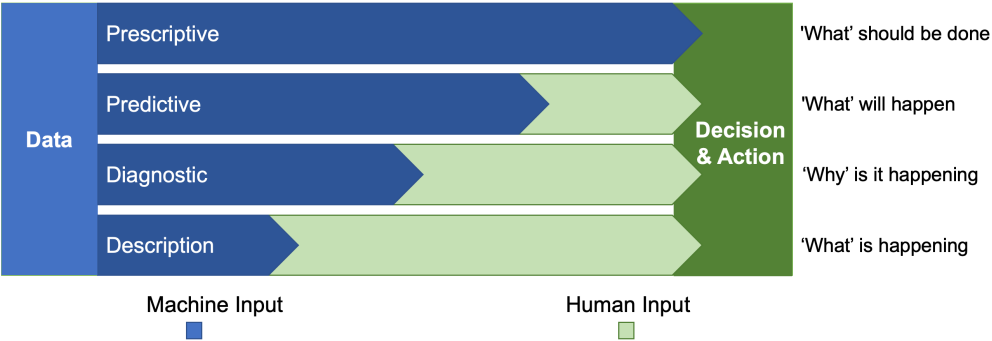

*Figure 2 Levels of analytics capabilities as suggested by Gartner*

## 2.3 Relevant works on selecting an appropriate level of DT

To the best of their knowledge, the authors are not aware of any existing work that focuses on selecting an appropriate level of DT. However, as the idea of selecting a level of DT is related to technology evaluation in general and the selection of technological capabilities for an organization in particular, the concepts discussed in these related fields are relevant and applicable to our research. Therefore, a summary of the existing literature is provided as a precursor to the digitalization framework.

For years, in the broader strategic management literature, scholars have been juxtaposing the forces that shape technology selection (Schmookler 2013; Myers et al. 1969; Rosenberg and Nathan 1982). Primarily, two major models have been proposed to guide the selection of appropriate technological capabilities (1) need pull and (2) technology push (Chau and Tam 2000; Nemet 2009; Horbach et al. 2012; Di Stefano et al. 2012). In the AEC literature, (Nam and Tatum 1992) also observed similar models guiding the technology selection process among the construction companies. The following paragraphs summarize these two prominent models for technology selection and detail how they relate to the digitalization framework.

The need pull model assumes that the problem (or the need) to be solved acts as the driving force for selecting the technological capabilities. Popular technology evaluation methods like comparing the internal rate of return, payback period, and strategic fit (Milis and Mercken 2004; Love et al. 2005; Stockdale et al. 2006) fall under this category. They all, in some form, select the technology based on the expected value, even if the precise definition of "value" varies across each method.

Although the need pull model might seem very intuitive, it is not suitable when the problem is not detected apriori, or the technology cannot solve the pre-defined problems. Moreover, the developments associated with big data and technology exaggerate this further. For example, much of the data collection today is not intentional (or planned to solve a problem) (Varian 2010). It is haphazard, heterogeneous, and messy. This data can sometimes uncover nontrivial insights about the problems which were not originally intended or planned. To alleviate this issue, the technology push model suggests technology acting as a lead instead of reacting to the business problem. It selects the existing technology capability for deployment and searches for an appropriate problem it can solve. However, this approach lacks critical evaluation of the business value (Love et al. 2020), and the corresponding changes required in the organizational conditions to sustain the value (Love and Matthews 2019), resulting in the inability to realize the purported benefits of the technology.

It is quite visible that both models have their pros and cons. Therefore, to alleviate the shortcomings, (Burgelman and Sayles 1988; Brem and Voigt 2009) suggest an appropriate alignment between the approaches. This idea of alignment between the need pull and technology push to select technological capability forms the central theme in the digitalization framework. More discussion on this is provided in Section 4.

## 2.4 Gap in knowledge

To enable a widespread adaption of DT in practice, there is a need to enable the practitioners select an appropriate level of sophistication in a DT. In addition, a framework/method is also needed which can allow the practitioners to create a strategic roadmap for the implementation of DT. The existing studies in the literature do not sufficiently address these issues.

A review of the literature reveals that although there exist several methods for general technology evaluation and selection, none of them have been used in the context of DT. Moreover, the prominent models for selecting general technological capabilities highlight a seeming dichotomy. On the one hand, the need pull model anchors on the problem and lacks the consideration of whether the technology can deliver the envisioned value. On the other hand, the technology push model ensures that the technological capability is available at the starting point but tends to miss the business value analysis.

To address the gap in knowledge, this work focuses on building the digitalization framework for the deployment of DT in the AEC industry. It considers both the need pull and technology push perspective and emphasizes the need to align both the approaches. The work further validates its finding through expert feedback following a Design Science Research methodology.

## 3. RESEARCH METHODOLOGY

The digitalization framework has been developed and validated over the course of 18 months by using the Design Science Research (DSR) or Constructive research methodology (Holmström et al. 2009). The DSR methodology allows for practical problem solving along with theoretical knowledge creation (Geerts 2011), which cannot be typically achieved by research methods such as surveys and questionnaires (AlSehaimi et al. 2013). It tends to focus on describing, explaining, and predicting the current natural or social world by not only understanding problems but also designing solutions to improve performance (Van Aken 2005). Specifically, the DSR methodology develops constructs (e.g., conceptual models, methods, frameworks, or artefacts) that are relevant in practice and, at the same time, ensure conceptual rigor. These constructs do not describe the existing reality like the approaches in explanatory sciences (e.g., sociology and natural sciences) but instead help create a new desired reality (Järvinen 2007).

As the digitalization framework is built to provide a rigorous conceptualization for selecting a level of DT, one that is relevant in practice, DSR methodology is a good fit. The DSR methodology has been used to develop several practical and technological artefacts in the construction industry. Oyegoke and Kiiras (2009) used the constructive approach to develop an innovative procurement method for improving owner's contracting strategies. Tezel et al. (2021) developed a blockchain model for supply chain management in construction. Chu et al. (2018) used the DSR methodology to develop a framework for integrating BIM and augmented reality in construction management.

Following the DSR methodology suggested by (Peffers et al. 2007), a five-stage process for the development of the digitalization framework was followed as shown in Figure 3: (1) problem identification, (2) defining research objectives, (3) framework design and development, (4) framework demonstration and evaluation, and (5) framework usefulness testing. The following paragraphs detail each of the stages.

In Stage 1, the authors identified the problem through self-reflection and an initial literature review as described by (Hevner and Chatterjee 2010). Practitioner's problem in selecting appropriate capabilities in a DT was observed during an ethnographic action research study, which was a part of the author's previous work (Agrawal et al. 2022). The existence of the problem was further validated through a literature review, as summarized in Section 2.

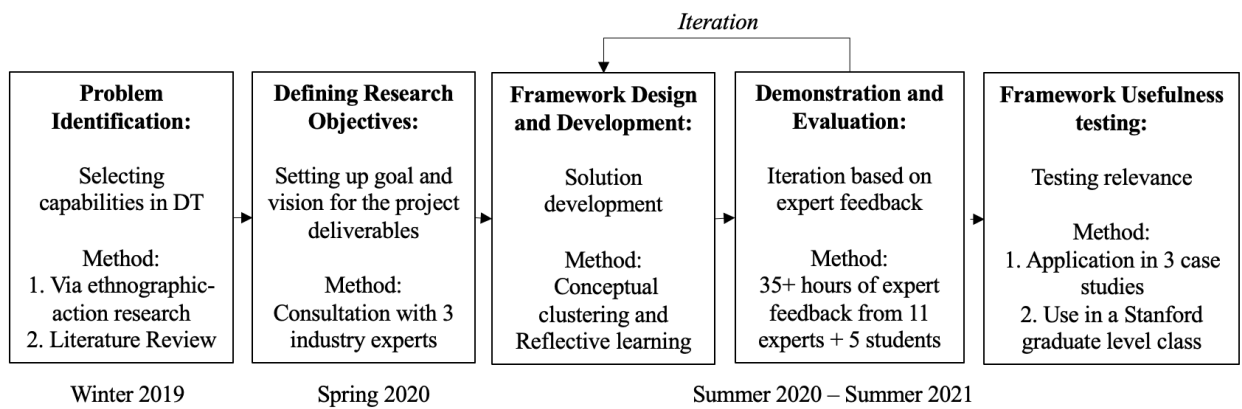

*Figure 3 Research Methodology*

Stage 2 involved defining specific research objectives for the study. The authors got together with three industry experts to formulate the vision for the project as: *"Develop a detailed and validated digital strategy framework to gain actionable insights,"* emphasizing the potential benefits of DT in the context of an organization or a specific project. The research objectives and the initial project proposal have been documented digitally (Agrawal and Fischer 2020).

Stage 3 (Framework design and development) and Stage 4 (Demonstration and evaluation) happened simultaneously in an iterative manner. The framework design and development followed a similar process as (Succar and Poirier 2020; Agrawal et al. 2021). The initial development of the framework was to understand how people selected technological capabilities in a DT. For this, the authors conducted an extensive literature review (described in Section 2). Following interactive field research and executive experience methodology, as described by (Burgelman and Siegel 2007; 2008), the authors supported the initial framework development with additional data obtained through the author's academic, executive, consulting, and broad experiences gained over many years.

The initial framework development aimed to answer questions like: What themes/factors affect DT selection? Do these factors have to support/oppose each other? Who is responsible for aligning these factors? The themes and concepts for the framework were identified using retroduction, conceptual clustering, and reflective learning (Shapiro 1992; Van Der Heijden and Eden 1998; Walker et al. 2008) based on the answers to the questions.

Once a preliminary version of the framework was prepared, it was demonstrated to experts (see Table 1) for feedback, and multiple design and development iteration were carried out. The experts were selected based on three factors: (1) practical experience and conceptual knowledge in the topics covered by the study (e.g., DT, technology adaption, technology strategy, and innovation management), (2) diversity in professional roles held by experts to ensure complementary skills and thinking, and (3) willingness of experts to get involved in multiple feedback sessions over next several months.

The demonstrations/feedback sessions were in the form of regular check-ins with the experts. 30+ meetings with 11 experts and 5 graduate students, clocking about 35 hours, were conducted during 18 months. The demonstration sessions with the experts were semi-structured and followed a similar protocol as used by (Agrawal et al. 2021). The session started with presenting the latest iteration of the framework to the expert and recording their feedback. The experts were asked to specifically comment on four parts: (1) elements of the framework that made sense, (2) elements of the framework that needed to be improved, (3) perceived helpfulness of the framework in practice, and (4) perceived comfort level in using the framework in practice.

The feedback from each session was incorporated into the next iteration. Multiple meetings with the same expert were conducted to ensure that the feedback was incorporated appropriately. These design iterations were carried out until theoretical saturation was reached, i.e., no new or relevant feedback emerged from the demonstrations (Glaser and Strauss 2017). To provide a trail for the development of the framework, three old but pivotal versions of the framework: digitalization pyramid, spring model, and the 3-force 3-factor model have been documented digitally (Agrawal and Fischer 2019; 2020).

Finally, in Stage 5, to analyze the practical usefulness, two studies were conducted:

1. The framework was applied along with practitioners to three real-life case studies to see if it can provide some actionable insights. More details on these case studies are in Section 5.
2. To further validate if the framework can be used independently, the digitalization framework was used in a three-month, graduate-level project-based class at Stanford University. The students (total 5 groups) applied the framework in their projects involving DT implementation for a commercial company. It was found that the framework is helpful to highlight the alignments and misalignments between the business and technology strategies. The framework was also found to be useful in setting informed management expectations upfront and creating a technology implementation plan. The course description, project description, and the class syllabus can be found digitally at (Fischer and Agrawal 2021).

*Table 1 Demonstration of the framework, profile of experts*

|  | Expert Code | Background/Role | Experience (Years) | Number of Meetings | Total hours of interaction |
|---|---|---|---|---|---|
| Industry Experts | A | Senior manager in a construction firm | 20-25 | 3 | 3 |
|  | B | Project manager on a $1B+ project | 25-30 | 1 | 2 |
|  | C | Senior manager in a construction firm | 15-20 | 2 | 1 |
|  | D | Head of innovation in AEC firm | 20-25 | 2 | 1 |
|  | E | Management executive in AEC firm | 30-35 | 1 | 1 |
|  | F | Project manager in a construction firm | 10-15 | 3 | 2 |
|  | G | Innovation lead with a general contractor | 10-15 | 1 | 1 |
|  | H | Researcher in use of AI in AEC industry | 30-35 | 5 | 5 |
|  | I | Expert and researcher in business strategy | 35-40 | 2 | 2 |
|  | J | Researcher in innovation management | 10-15 | 3 | 3 |
|  | K | Researcher in technology deployment | 20-25 | 2 | 2 |
| Graduate Students | S1 | Researcher in deployment of ML in AEC industry | 0-5 | 3 | 3 |
|  | S2 | Researcher in Virtual design and construction | 5-10 | 5 | 5 |
|  | S3 | ML researcher in AEC industry | 0-5 | 1 | 1 |
|  | S4 | Researcher in industrial facilities management | 0-5 | 1 | 2 |
|  | S5 | Researcher in building operations management | 0-5 | 1 | 1 |

*Total of 23-hours of interaction with 11 experts and 12-hours interaction with 5 graduate students spreading over 18-months*

# 4. DIGITALIZATION FRAMEWORK

The digitalization framework is an integration of two different perspectives (see Figure 4) about technology deployment existing in the strategic management literature, as described below:

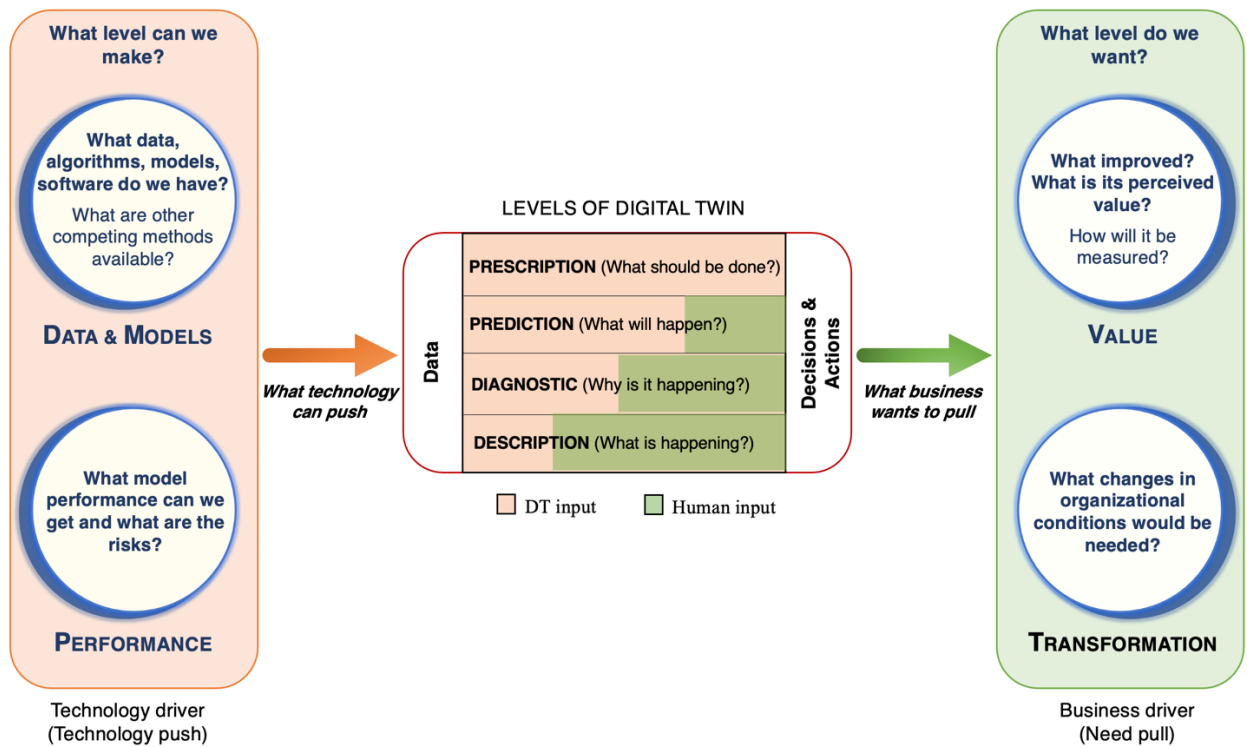

*Figure 4 The digitalization framework*

*Need pull perspective:* This is anchored on the notion that business needs, where a company's management wants to achieve a particular value or a competitive advantage, are the reason for deploying a DT. Therefore, in these situations, the problem (or the need) that is to be solved through a DT is clear and acts as the driving force for selecting the appropriate level of DT and the technological capabilities needed to support it. For example, consider a case where a firm is looking for a way to improve highway maintenance because the current process of managing the highway is very inefficient. In this case, the driver for the deployment of DT is the need to improve the current maintenance process of the highway. The firm would try to figure out the top reasons for this inefficiency and build a DT that can solve it.

*Technology push perspective:* Describes technology as a driver for deploying new solutions. In these situations, a "technology champion" or any other person in the company, fascinated by technology developments decides to change something. This scenario may not be necessarily motivated by a pressing "need" in the first place and might require searching for an appropriate use case for deploying the technology. For example, a firm can be fascinated by the recent developments in ML and AI and search for some use cases to improve the current business.

In essence, the need pull perspective emphasizes the business strategy informing the technology deployment, and the technology push perspective emphasizes the need to change the business strategy according to the evolving technological capabilities. We feel that both perspectives are equally important. A DT deployment motivated by a problem in hand, and lacking the technological capabilities to build it, will lead to unrealistic expectations from the technology and ultimately failure and frustration. On the other hand, deploying a DT just because of the technology fascination can result in wastage of resources and unrealized benefits from DT.

Therefore, the digitalization framework emphasizes the need to align both these approaches for a successful deployment of DT in practice, as highlighted by (Burgelman and Sayles 1988; Brem and Voigt 2009). Although, in practice, the main driver for the deployment of a DT can be any of these perspectives, it is essential to keep the other part in mind. Hence, the framework emphasizes the *'level which the technology can push'* and the *'level which the business wants to pull'*.

Figure 4 shows the two perspectives in the context of levels of DT, with the technology driver on the left and the business driver on the right. The business driver anchors on the elements: (1) Value and (2) Transformation, and the technology driver anchors on the elements: (1) Data & Models and (2) Performance. Each of these elements has been explained below. The central element, levels of DT, is generalizable, and any DT hierarchy can be used. For this paper, hierarchy inspired by Gartner is used, as explained in Section 2.

**Value:** In essence, the purpose of a DT is to improve the current ways of working in some form or the other. Therefore, it becomes important to ask: *"What has (or will be) improved through the use of a DT, and what value does that improvement bring to our business?"* 'Value' defines the extent of impacts that the organization wants or expects from using a technology (Renkema 2000). It also anchors the firm's business strategy, ensuring that the correct problem is being solved and thus drives the design of technological solutions. Lack of value evaluation of the technology confounds the organizations to realize the purported benefits of digitalization (Love and Matthews 2019).

Many construction organizations do not evaluate technology for value creation, and those that do, use very broad qualitative value definitions like: "improved growth and success" and "improved market share" (Love et al. 2004). The defined value should be quantified, measurable, and directly affected by the technology implemented. Defining the value too narrowly overlooks commonalities and linkages across different aspects of the product, the process, and the organization delivering it. On the other hand, an unquantified, broadly defined value makes the definition poor and vague, leading to no mechanism to remind the team of goals once work has begun (Fischer et al. 2017).

**Transformation:** Value cannot be delivered without a change (Peppard 2016). Benefits of technology deployment are marginal if only superimposed on the existing organizational conditions (Venkatraman 1994). Love and Matthews (2019) also notes that to realize the purported benefits of technology, asking a series of 'how' questions is essential. Therefore, to accrue the maximum benefits from DT deployment, it becomes important to ask: *"What changes in organization conditions would be needed?"* Agrawal et al. (2022) reports from an ethnographic action study that the magnitude of the changes required in the organizational processes to sustain the value generated by DT commensurate the technological capabilities that a DT possesses. A higher level of DT can lead to radical changes in an organization as compared to a lower level of DT. An example of this has been provided in Section-5.

**Data & Models:** Real-world problems are complex with a lot of unknowns. The end product we want is a piece of code (and possibly some hardware) that can solve the problem. But there is a huge chasm between the problem and the solution. Algorithms and models translate real-world problems into mathematical objects that the computer can understand and work upon. Thus, the models and algorithms, which in turn depend on the data and programming abilities in hand, to a large extent determine the types and extent of cognitive ability in a DT.

To ensure appropriate technological capability for deploying a DT, it becomes important to ask: *"What data, algorithms, models, software do we have?"* Generally, modeling is lossy and cannot capture all the richness of the real world. British statistician George Box famously quotes: "All models are wrong, but some are useful," indicating that the model will never reflect the exact real-world behavior. This makes it important to ensure that the model explicitly captures the part of the real world that is important for answering the user's question, again highlighting the need for an alignment between the two perspectives.

**Performance:** The performance measure is what we would evaluate a DT against and would like our DT to aspire for. In other words, it is the point at which we can say that the model (and algorithm) is good enough for deployment. It includes all the desirable qualities like having high accuracy, performing the task in minimum possible time, and requiring minimum human effort. Some of these goals would conflict, and a trade-off would be necessary. Thus, it is essential to estimate the required performance levels for each goal and set the relative importance between them based on the risks involved. From practice, we have observed that practitioners tend to:

1. *Confuse Value and Performance measures*: We have often seen practitioners defining the performance requirements for a DT as "reducing the cost of the project," "speeding up the decision latency," and "increasing safety on the site." These sample excerpts by the practitioners are the value that can be derived from a DT and not the performance measure against which DT would be evaluated. Sample performance measures in these cases would be: "60% faster schedule creation" (leads to the value: "reducing the cost of the project"), and "automated decisions within 30 minutes of a query with at least 90% accuracy" (leads to the value: "speeding up decision latency").

2. *Set up qualitative performance requirements*: We have also observed practitioners setting up requirements like "high accuracy," "good level of accuracy," and "within a reasonable time." Such qualitative requirements are impossible for the model to interpret and thus become a hindrance in building a model that can be deployed in practice. We suggest that the defined performance requirements be interpretable by any randomly selected person without using their judgment. For example, a performance requirement defined as "good accuracy" is subject to the judgment of the person of what it means to be "good," but a requirement defined as "at least 90% accurate" is objective and does not require any judgment.

After introducing the digitalization framework, it is important to showcase its applications in real-world case studies. Section 5 presents three real-life case studies using the digitalization framework. Section 6 follows by expanding on the findings and additional use cases of the framework.

## 5. APPLICATION OF THE FRAMEWORK IN CASE STUDIES

Perhaps the most appropriate test of the usefulness of a conceptual framework is whether practitioners in companies find value in applying it in practice. Therefore, to bring the framework to life, we demonstrate the framework's application in three real-life case studies. Though its application in three case studies cannot claim the generality of the framework's usefulness, the authors have tried to demonstrate the application across different project phases with different types of AEC firms situated in various parts of the world.

The first case study is of construction engineering company and was retrospective, trying to identify if the framework could have partly helped resolve the company's issues during the project. The second case study is of a full-service engineering firm and was prospective, where the framework was applied in the planning/pilot phase to help the project team better understand the plausible issues that could arise while deploying a DT. The third case study is of a general contractor and was applied in real-time during the project and helped solve some of the team's issues.

**Case Study-1: DT deployment on a wind turbine project**

*Situation Description:*

The case study focuses on a construction firm involved in the project of installing and maintaining wind turbines. One of the cranes installing the wind turbines on the project recently experienced a tip over due to soil failure. Crane failures are extremely dangerous and costly for the project. The project team thus realized the importance of calculating the soil bearing capacity before walking the crane. They estimated that the manual method of data collection, data processing, and calculation takes around eight weeks to get the results, which was unacceptable. Thus, the team envisioned creating a DT of the installation site, which can have the real-time data of soil and topography and predict the soil capacity instantaneously via Machine Learning (ML).

*Key events:*

1. Project manager (R) selects the level of DT as "*Prediction*" and envisions using a DT to collect real-time soil data and predict the soil capacity using ML.
2. R defines the "*Value*" from DT as improving the soil capacity calculation time from eight weeks to real-time via prediction from the ML model, which would help the company to have zero crane tip-overs.

3. R checks with the senior management regarding the *"Transformation"* and decides to (1) hire a data science team for building the model and (2) change some of the current decision-making policies in the company regarding the crane walk, which were based on manual observations rather than data-backed calculations.

4. The data team builds a model achieving an accuracy of 90%, which R terms as unacceptable. The data team tells R that getting to a higher accuracy would require more data and sophisticated models that the company did not have at that time.

5. R realizes that they do not have the technological capability to support the *"Prediction"* level. Thus, R decides to move to a lower level of DT - *"Description"* - where DT can help collect and pre-process the data. The calculations would still be done manually. Though the team could not get instantaneous results, this helped reduce the time from 8 weeks to 3 days.

*Observations and Insights:*

Even though the project team could finally reach an appropriate level of DT, the whole iterative process took them around three months. The main issue was uninformed expectations from the technology, where the team lacked clarity about what the technology could and could not do. ML was selected partly due to technology fascination rather than following a rational selection approach. One of the members of the data science team stated: "If R had told us at the start of the project that they were looking for accuracy of over 99.9%, I would have said upfront that this would be very difficult to get with the amount of data we have."

Retrospective analysis with R revealed that the project team did not formulate the *"Performance"* element of the framework or define what a "good" model means for their scenario. R further added that the team was utterly confused about what a good model might mean and that different people had a different opinions about it. Due to a lack of formulation of the performance element, the project team could not anticipate what DT could deliver, leading to unrealistic expectations from the DT. On the other hand, the data science team did not have an objective way to evaluate their ML models. They went off their experience to anticipate that 90% should be a good enough accuracy, failing to comprehend the gravity of the situation: *a 90% accuracy means that there is a chance 10 out of 100 cranes can fail*.

From our experience, we have observed this is a common scenario where the project team is indecisive and unable to articulate the performance they want from a DT. This often leads to misalignments between what technology can deliver and what the business wants, as demonstrated in this case study. It is important to note that even though all other elements of the framework were relatively well defined in this case study, the lack of one element resulted in the misalignment, highlighting the importance and the need to pay attention to all the aspects of the framework.

As this was a retrospective case study, we cannot test if the situation would have been different if the digitalization framework had been used. However, R did acknowledge that possibly using the digitalization framework at the start of the project could have helped the project team align the expectations from DT. This would have enabled a better selection of the appropriate level of DT and possibly reduced the three months turnaround time.

**Case Study-2: DT deployment on a highway maintenance project**

*Situation Description:*

The case study focuses on a $1.35 billion publicly owned toll road (SH-130 in Austin, Texas) stretching over 41 miles. The toll road was opened in 2012 and is operated and maintained under the terms of a 50-year facility concession agreement with the Texas Department of Transportation (TxDoT). The performance requirements indicating the minimum expected performance levels and the defect resolution times were set upfront in the SH-130 concession agreement by the Department of Transportation (DoT). Some requirements necessitated a very stringent response time - 6 to 24 hours. Failing to respond adequately could result in heavy fines and risks to the life safety of the users. The manager (B) on this project wondered if a DT could help in some ways.

*Key events:*

1. B envisions creating a DT of the roadway. As the framework was applied in the planning phase, B did not have an apriori level of DT in mind and was open to brainstorming and suggestions.
2. B defines the envisioned *"Value"* from DT as reducing operational costs through early detection and preventative maintenance. Detecting the defects early would give the firm more time to resolve them, thus helping them achieve a lower operating cost and move towards a preventive maintenance system.
3. B identifies that early detection of defects could be achieved by the *"Description"* level of DT. A computer vision algorithm equipped in a DT can help it detect defects from the real-time data collected via drone imagery of the roadway.
4. B then defines the technological capabilities needed to build this DT and finds that the DT seems feasible with the capabilities that the firm currently possesses.

5. For the preventive maintenance system, B wonders if a DT can also help to predict the potential occurrence of cracks/defects. They check the technological capabilities needed and find them to be appropriate for their requirements.
6. When B moves on to formulate the *"Transformation"* needed at the *"Prediction"* level of the DT, some non-trivial insights emerge.

*Observations and Insights:*

While applying the digitalization framework during the planning phase of DT deployment, B observed that DT had a significant impact on the corresponding organizational conditions. A summary of the insights is presented below.

Figure 5 (a) shows the process followed by the firm for highway maintenance. The process starts by inspecting the elements and reporting any observed defects during the patrolling and inspections. This is followed by work order creation, checking of required budget and inventory, and designating the appropriate crew for pothole repair and inspection. As shown in the figure, humans are not so efficient in detecting pavement defects.

The *'Description'* level of DT only informs about the current situation of the world. This function was previously being performed by the highway patrol team consisting of engineers and the maintenance staff. With the deployment of DT, the process of detecting the defects, creating work orders, and inspecting the defects can be automated. DT would be able to detect the defects and understand the changes that happen over time. Therefore, B would need to make small changes in the organization and shift the personnel dedicated to these tasks earlier to more productive tasks, as shown in Figure 5 (b).

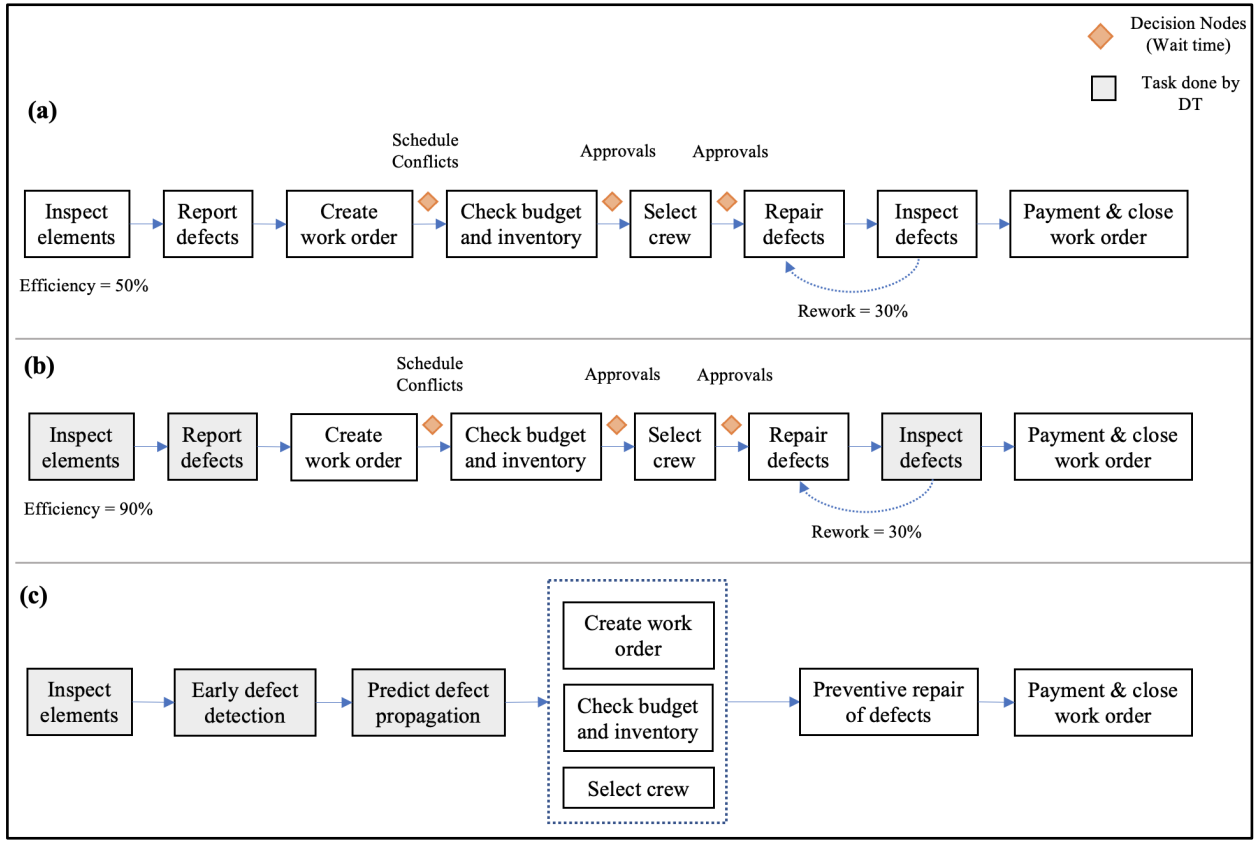

*Figure 5 (a) Original workflow, (b) Changes in workflow with DT at "Description" level; (c) New workflow with DT at "Prediction" level*

On the other hand, when the company moves to the *'Prediction'* level of DT, many process changes are required, as shown in Figure 5 (c). Due to the strict guidelines, the SH-130 concession company needs to keep a large amount of inventory and contact individual subcontractors in a short period of time to carry out the repair work, leading to unnecessary costs for the company. With DT making predictions, the team can plan for the material, crew, and equipment in advance. This modifies several of the steps in the workflow and allows them to be completed in parallel instead of doing them sequentially (see Figure 5 (c)). For example, on January 1st of every year, the company can query the defects that are expected to appear and create advanced work orders, thereby removing the latency and decision-making time. The company can also pivot to a "Just-in-time" delivery method for materials and equipment delivery. But it should be noted, moving to this kind of system would require significant remodeling of the current ways of working like changing the existing contracts, legal tenders, and business partnerships, again emphasizing the significance of *"Transformation"* to sustain the value provided by DT.

In a semi-structured interview, B acknowledged that the digitalization framework acted as a structured planning tool, forcing the project team to think through the process of DT deployment and articulate the transformations needed in the organization. B also added that the framework enabled him clearly communicate his strategic vision for DT deployment to the top management along with the corresponding changes needed in the organization. A follow-up interview with an executive from the top management of the company revealed that the framework helped them achieve clarity on DT deployment and made them much more confident about the project roadmap ahead. Ultimately, a successful deployment of DT would depend on resolution of these and many other factors.

**Case Study-3: DT deployment on a construction project**

*Situation Description:*

The first two case studies were driven by the need pull mechanism, with the project team looking to solve a specific issue using a DT. On the other hand, this case is driven by the technology push mechanism, with the company's management fascinated by DT and looking for ways it can add value to the business.

An innovation manager (G) working with a mid-sized general contractor in Norway develops a DT prototype for weekly/daily construction site management combining the BIM and the Last Planner System. The system worked very well on a few demonstration projects allowing foremen and the superintendent to understand the work accomplished throughout the day. G expected, therefore, rapid deployment of the prototype across many projects. This did not happen. Working through the framework, at least in part, helped G understand the issue behind this.

*Observations and Insights:*

Working through the framework from the technology perspective, it was realized that DT developed by G improved the *"Description"* of what is happening on site. Some site staff found this improvement helpful. Working through the diagram from the value side with G, it was realized that most site staff really wanted a "*Prescription,*" i.e., they wanted to know what should be done tomorrow and beyond. Hence, the insights expected by the business or value perspective were misaligned with the insights the technology could provide, being the major reason for the slow adaption of the technology.

In this case study, the digitalization framework was used as a diagnostic tool to detect G's problem in real-time while deploying DT. The framework helped G understand the misalignment between what the business wanted and what the technology was delivering. Once G understood the root cause of the problem, he decided to clearly communicate the benefits that the DT can provide to the management and thus align the expectations. G also realized the need for the *"Prescription"* level, as suggested by the staff members, and decided to move to it in the future by using ML.

## 6. CONCLUSION

For a successful deployment of DT, the managers and practitioners should ideally be able to select an appropriate level of sophistication in a DT, articulate the technological requirements to build it, and clearly communicate the strategic vision for its implementation to the top management. But given the varied range of capabilities that DT offers, practitioners themselves are confused and face increasingly difficult decisions regarding what type of technological capabilities to select in a DT while deploying it in the AEC industry. This confusion results in unrealistic hopes from the technology, strategic misalignments, and misallocation of resources, ultimately leading to a slower adaption of digital technologies in the AEC industry.

Therefore, to alleviate this confusion, the paper presents the digitalization framework that helps practitioners strategically select an appropriate level of sophistication in a DT. The framework brings together ideas from the strategic management literature emphasizing the importance of technology-driven problem-solving (technology push) with ideas that emphasize the importance of a value-driven approach for problem-solving (need pull), and argues that the alignment of both approaches is necessary for a successful deployment of DT. The framework has been developed and validated following the DSR research methodology over 18 months. The design iterations and the validation was carried out using the feedback from 11 experts and five graduate students over multiple meetings clocking over 35 hours. The framework has also been further tested for usefulness by: (1) applying it in three longitudinal case studies and (2) using it in five student projects in a graduate-level class at Stanford University. It should be noted that the current version of the digitalization framework has only been validated and tested through case studies and expert feedback in the AEC domain. Therefore, the generality and applicability of this framework cannot be claimed outside the AEC industry. Future research can explore the universality of this framework by conducting case studies and expert reviews from other industries as well.

The digitalization framework is intended to be useful for both academics and practitioners. For academics, the framework should help researchers better evaluate their proposed DT applications and models for the business value it provides, technological capabilities needed to build it, and the organizational changes required to sustain the value generated from DT. Awareness of these factors would enable the researchers to develop models and methods that are more likely to succeed in practice. The framework should also act as a gentle reminder for educators that there is probably less value in defining/envisioning a so-called most ideal or technologically advanced DT. The DT needs to be evaluated for each use case, and that a "one size fits all" approach would not work.

The practitioners should use the digitalization framework as a planning and diagnostic tool to objectively evaluate and understand the different forces in play while deploying a DT in practice and not to prescribe or declare a level of DT for a particular situation. Specifically, we envision the following three use cases for the framework in practice:

**Selecting an appropriate level of DT:** The digitalization framework enables the practitioners to systematically examine different levels of DT and choose the one that provides the most value in the light of existing technological capabilities. It helps them articulate what DT would provide (business value) and understand what it would take to get there (technological capabilities). Jahanger et al. (2021) identifies the lack of knowledge regarding the challenges and barriers in digital implementations among owners and contracts as one of the main factors for the lag in adapting digital technology. Therefore, using the digitalization framework as a planning tool, especially at the start of a project and at pivotal intermediate points, helps in clear communication of the goals and thus create a shared understanding among the participants of what to expect and what resources to allocate, as shown in case study 1.

**Better stakeholder management and formulating a digital strategy:** Reaching an "ideal" envisioned DT is not a one-shot task. There can be instances of misalignments between what the business expects and what the technology can deliver. Such misalignments can often lead to delayed DT deployment, if not whole project's dissolution. Reviewing technology implementations in the AEC industry over 15 years, Y. Lu et al. (2015) reports that a common reason for unsuccessful technology investments was lack of necessary support from project clients and professional consultants, emphasizing the need of a strategic alignment among different stakeholders for a successful technological deployment. In such critical situations, the digitalization framework can act as a diagnostic tool by helping highlight and understand the root cause for these misalignments and therefore start a conversation towards its resolution, as shown in case study 3.

Suppose the business wants to pull a higher level of DT, and the technology cannot push it. In that case the management has three choices: (1) change their expectations to accept the existing level of DT, (2) invest into building a higher level of DT that can fulfil their expectations, or (3) scrap the project altogether. On the other hand, if the business wants to pull a lower level of DT than what the technology can push, like people using BIM only for documentation/representational purposes even though it can offer much more, the management again has three choices: (1) keep using DT at a lower level, (2) find use cases which can make full use of DT, or (3) change the current DT to only have the required capabilities. In these scenarios, the digitalization framework helps the managers formulate a long-term vision of which level of DT to reach and a strategic roadmap for reaching there from the current level of DT.

**Inculcating a strategic mindset throughout the organization:** Owing to the different areas of expertise in a company, some people might be closer to the technology without understanding the business value and vice-versa, leading to misalignment in organizational strategies. Ansari et al. (2015) notes that many challenges emerge due to the strategies made by top-level managers not having good effects on the operational level of the organization. Therefore, to avoid missed opportunities, it becomes essential to have a healthy and open debate among the company's executives of different ranks about the evolving technological landscape and its implication for the business.

The digitalization framework facilitates such strategic conversations, as shown in case study 2. The jargon-free nature of the framework makes it easy to communicate and internalize its basic lessons (Burgelman and Siegel 2008). Even the executives who are missing the background in either the business strategy or the technical capabilities can reason and understand the other side. Therefore, the natural language form of the digitalization framework can become a useful tool to inculcate a strategic mentality throughout the organization and stimulate continuous technological discussions.

Ultimately, there is no one universal answer to the question: "What capabilities should be selected in a DT for deployment?" The management of the company must continue to carefully evaluate all the possibilities and select the level of sophistication which provides the maximum value with the technological abilities in hand. Business expecting a higher level of sophistication in a DT, and the technology not being able to deliver, can result in false hopes and ultimately rejection of DT as hype. On the other hand, technology delivering a highly sophisticated DT and the business not appreciating its value, can result in missed opportunities, and unrealized benefits from the technology. The digitalization framework facilitates the process of selection by forcing the practitioners to articulate the perceived business value and the technological capabilities needed in the DT. Awareness of these factors across the firm can increase the likelihood of a successful deployment and adaption of a DT.

## 7. DATA AVAILABILITY STATEMENT

Some or all data, models, or code generated or used during the study are proprietary or confidential in nature and may only be provided with restrictions.

## 8. ACKNOWLEDGEMENTS

We would like to thank all the experts who gave us their valuable time and feedback. We also thank the graduate students who used the framework in their class and made this study stronger. We acknowledge the financial support provided by Center for Integrated Facility Engineering (CIFE) at Stanford University. The authors would also take this opportunity to express their gratitude to Tulika Majumdar, Rui Liu, Hesam Hamledari, and Alberto Tono for their valuable feedback on the framework and the manuscript.## 9. REFERENCES

Agrawal, Ashwin, and Martin Fischer. 2019. "Digital Twin for Construction | Center for Integrated Facility Engineering." 2019. https://cife.stanford.edu/Seed2019%20DigitalTwin.
———. 2020. "Digital Strategy for Construction | Center for Integrated Facility Engineering." 2020. https://cife.stanford.edu/Seed20digital-strategy-construction.
Agrawal, Ashwin, Vishal Singh, Robert Thiel, Michael Pillsbury, Harrison Knoll, Jay Puckett, and Martin Fischer. 2022. "Digital Twin in Practice: Emergent Insights from an Ethnographic-Action Research Study." *ASCE Construction Research Congress 2022*.
Agrawal, Ashwin, Robert Thiel, Pooja Jain, Vishal Singh, and Martin Fischer. 2021. "Levels of Digital Twin." *(Submitted for Review)*.
Akula, Manu, Robert R. Lipman, Marek Franaszek, Kamel S. Saidi, Geraldine S. Cheok, and Vineet R. Kamat. 2013. "Real-Time Drill Monitoring and Control Using Building Information Models Augmented with 3D Imaging Data." *Automation in Construction* 36 (December): 1–15. https://doi.org/10.1016/j.autcon.2013.08.010.
AlSehaimi, Abdullah, Lauri Koskela, and Patricia Tzortzopoulos. 2013. "Need for Alternative Research Approaches in Construction Management: Case of Delay Studies." *Journal of Management in Engineering* 29 (4): 407–13. https://doi.org/10.1061/(ASCE)ME.1943-5479.0000148.
Al-Sehrawy, Ramy, and Bimal Kumar. 2021. "Digital Twins in Architecture, Engineering, Construction and Operations. A Brief Review and Analysis." In *Proceedings of the 18th International Conference on Computing in Civil and Building Engineering*, edited by Eduardo Toledo Santos and Sergio Scheer, 924–39. Lecture Notes in Civil Engineering. Cham: Springer International Publishing. https://doi.org/10.1007/978-3-030-51295-8_64.
Ansari, Ramin, Eghbal Shakeri, and Ali Raddadi. 2015. "Framework for Aligning Project Management with Organizational Strategies." *Journal of Management in Engineering* 31 (4): 04014050. https://doi.org/10.1061/(ASCE)ME.1943-5479.0000249.
Austin, Mark, Parastoo Delgoshaei, Maria Coelho, and Mohammad Heidarinejad. 2020. "Architecting Smart City Digital Twins: Combined Semantic Model and Machine Learning Approach." *Journal of Management in Engineering* 36 (4): 04020026. https://doi.org/10.1061/(ASCE)ME.1943-5479.0000774.
Autodesk. 2021. "Digital Twins in Construction, Engineering, & Architecture | Autodesk." 2021. https://www.autodesk.com/solutions/digital-twin/architecture-engineering-construction.
Boje, Calin, Annie Guerriero, Sylvain Kubicki, and Yacine Rezgui. 2020. "Towards a Semantic Construction Digital Twin: Directions for Future Research." *Automation in Construction* 114 (June): 103179. https://doi.org/10.1016/j.autcon.2020.103179.
Boschert, Stefan, and Roland Rosen. 2016. "Digital Twin—The Simulation Aspect." In *Mechatronic Futures: Challenges and Solutions for Mechatronic Systems and Their Designers*, edited by Peter Hehenberger and David Bradley, 59–74. Cham: Springer International Publishing. https://doi.org/10.1007/978-3-319-32156-1_5.38

Brem, Alexander, and Kai-Ingo Voigt. 2009. "Integration of Market Pull and Technology Push in the Corporate Front End and Innovation Management—Insights from the German Software Industry." *Technovation*, Technology Management in the Service Economy, 29 (5): 351–67. https://doi.org/10.1016/j.technovation.2008.06.003.

Bueno, Martín, Frédéric Bosché, Higinio González-Jorge, Joaquín Martínez-Sánchez, and Pedro Arias. 2018. "4-Plane Congruent Sets for Automatic Registration of as-Is 3D Point Clouds with 3D BIM Models." *Automation in Construction* 89 (May): 120–34. https://doi.org/10.1016/j.autcon.2018.01.014.

Burgelman, Robert A., and Leonard R. Sayles. 1988. *Inside Corporate Innovation*. Simon and Schuster.

Burgelman, Robert A., and Robert E. Siegel. 2007. "Defining the Minimum Winning Game in High-Technology Ventures." *California Management Review* 49 (3): 6–26. https://doi.org/10.2307/41166392.

———. 2008. "Cutting the Strategy Diamond in High-Technology Ventures." *California Management Review* 50 (3): 140–67. https://doi.org/10.2307/41166449.

Canedo, Arquimedes. 2016. "Industrial IoT Lifecycle via Digital Twins." In *2016 International Conference on Hardware/Software Codesign and System Synthesis (CODES+ISSS)*, 1–1.

Chau, P. Y. K., and K. Y. Tam. 2000. "Organizational Adoption of Open Systems: A 'Technology-Push, Need-Pull' Perspective." *Information & Management* 37 (5): 229–39. https://doi.org/10.1016/S0378-7206(99)00050-6.

Chu, Michael, Jane Matthews, and Peter E. D. Love. 2018. "Integrating Mobile Building Information Modelling and Augmented Reality Systems: An Experimental Study." *Automation in Construction* 85 (January): 305–16. https://doi.org/10.1016/j.autcon.2017.10.032.

Cimino, Chiara, Elisa Negri, and Luca Fumagalli. 2019. "Review of Digital Twin Applications in Manufacturing." *Computers in Industry* 113 (December): 103130. https://doi.org/10.1016/j.compind.2019.103130.

Davenport, Thomas, and Jeanne Harris. 2017. *Competing on Analytics: Updated, with a New Introduction: The New Science of Winning*. Harvard Business Press.

Di Stefano, Giada, Alfonso Gambardella, and Gianmario Verona. 2012. "Technology Push and Demand Pull Perspectives in Innovation Studies: Current Findings and Future Research Directions." *Research Policy* 41 (8): 1283–95. https://doi.org/10.1016/j.respol.2012.03.021.

Du, Jing, Qi Zhu, Yangming Shi, Qi Wang, Yingzi Lin, and Daniel Zhao. 2020. "Cognition Digital Twins for Personalized Information Systems of Smart Cities: Proof of Concept." *Journal of Management in Engineering* 36 (2): 04019052. https://doi.org/10.1061/(ASCE)ME.1943-5479.0000740.

Ezhilarasu, Cordelia Mattuvarkuzhali, Zakwan Skaf, and Ian K Jennions. 2019. "Understanding the Role of a Digital Twin in Integrated Vehicle Health Management (IVHM)*." In *2019 IEEE International Conference on Systems, Man and Cybernetics (SMC)*, 1484–91. https://doi.org/10.1109/SMC.2019.8914244.

Fan, Chao, Yucheng Jiang, and Ali Mostafavi. 2020. "Social Sensing in Disaster City Digital Twin: Integrated Textual–Visual–Geo Framework for Situational Awareness during Built Environment Disruptions." *Journal of Management in Engineering* 36 (3): 04020002. https://doi.org/10.1061/(ASCE)ME.1943-5479.0000745.

Feng, B., S. Kim, S. Lazarova-Molnar, Z. Zheng, T. Roeder, and R. Thiesing. 2020. "A Case Study of Digital Twin for Manufacturing Process Invoving Human Interaction." 2020. https://www.semanticscholar.org/paper/A-CASE-STUDY-OF-DIGITAL-TWIN-FOR-A-MANUFACTURING-Feng-Kim/50f3cd21e4860470e8d762fb591193868d9fe878.

Fischer, Martin, and Ashwin Agrawal. 2021. "CEE-329 Class, Stanford University Syllabus." 2021. https://docs.google.com/document/d/1xfqhuQPV48aRYp6TyaPsnrgHxwx6I9EOVZgeV9Wi_4o/edit

Fischer, Martin, Howard W. Ashcraft, Dean Reed, and Atul Khanzode. 2017. *Integrating Project Delivery*. John Wiley & Sons.

Ford, David N., and Charles M. Wolf. 2020. "Smart Cities with Digital Twin Systems for Disaster Management." *Journal of Management in Engineering* 36 (4): 04020027. https://doi.org/10.1061/(ASCE)ME.1943-5479.0000779.

Francisco, Abigail, Neda Mohammadi, and John E. Taylor. 2020. "Smart City Digital Twin–Enabled Energy Management: Toward Real-Time Urban Building Energy Benchmarking." *Journal of Management in Engineering* 36 (2): 04019045. https://doi.org/10.1061/(ASCE)ME.1943-5479.0000741.

Gabor, Thomas, Lenz Belzner, Marie Kiermeier, Michael Till Beck, and Alexander Neitz. 2016. "A Simulation-Based Architecture for Smart Cyber-Physical Systems." In *2016 IEEE International Conference on Autonomic Computing (ICAC)*, 374–79. https://doi.org/10.1109/ICAC.2016.29.

Gartner. 2013. "Extend Your Portfolio of Analytics Capabilities." Gartner. 2013. https://www.gartner.com/en/documents/2594822/extend-your-portfolio-of-analytics-capabilities.

———. 2019. "Gartner Survey Reveals Digital Twins Are Entering Mainstream Use." Gartner. 2019. https://www.gartner.com/en/newsroom/press-releases/2019-02-20-gartner-survey-reveals-digital-twins-are-entering-mai.

Geerts, Guido L. 2011. "A Design Science Research Methodology and Its Application to Accounting Information Systems Research." *International Journal of Accounting Information Systems*, Special Issue on Methodologies in AIS Research, 12 (2): 142–51. https://doi.org/10.1016/j.accinf.2011.02.004.

Glaessgen, Edward, and David Stargel. 2012. "The Digital Twin Paradigm for Future NASA and U.S. Air Force Vehicles." In *53rd AIAA/ASME/ASCE/AHS/ASC Structures, Structural Dynamics and Materials Conference<BR>20th AIAA/ASME/AHS Adaptive Structures Conference<BR>14th AIAA*. Honolulu, Hawaii: American Institute of Aeronautics and Astronautics. https://doi.org/10.2514/6.2012-1818.

Glaser, Barney G., and Anselm L. Strauss. 2017. *Discovery of Grounded Theory: Strategies for Qualitative Research*. Routledge.

Grieves, Michael, and John Vickers. 2017. "Digital Twin: Mitigating Unpredictable, Undesirable Emergent Behavior in Complex Systems." In *Transdisciplinary Perspectives on Complex Systems: New Findings and Approaches*, edited by Franz-Josef Kahlen, Shannon Flumerfelt, and Anabela Alves, 85–113. Cham: Springer International Publishing. https://doi.org/10.1007/978-3-319-38756-7_4.

Ham, Youngjib, and Jaeyoon Kim. 2020. "Participatory Sensing and Digital Twin City: Updating Virtual City Models for Enhanced Risk-Informed Decision-Making." *Journal of Management in Engineering* 36 (3): 04020005. https://doi.org/10.1061/(ASCE)ME.1943-5479.0000748.

Hampson, Keith D., and C. B. Tatum. 1993. "Technology Strategy for Construction Automation." *Automation and Robotics in Construction X*, 125–33.

Hevner, Alan, and Samir Chatterjee. 2010. "Design Science Research in Information Systems." In *Design Research in Information Systems: Theory and Practice*, edited by Alan Hevner and Samir Chatterjee, 9–22. Integrated Series in Information Systems. Boston, MA: Springer US. https://doi.org/10.1007/978-1-4419-5653-8_2.

Holmström, Jan, Mikko Ketokivi, and Ari-Pekka Hameri. 2009. "Bridging Practice and Theory: A Design Science Approach." *Decision Sciences* 40 (1): 65–87. https://doi.org/10.1111/j.1540-5915.2008.00221.x.

Horbach, Jens, Christian Rammer, and Klaus Rennings. 2012. "Determinants of Eco-Innovations by Type of Environmental Impact — The Role of Regulatory Push/Pull, Technology Push and Market Pull." *Ecological Economics* 78 (June): 112–22. https://doi.org/10.1016/j.ecolecon.2012.04.005.

Jahanger, Qais K., Joseph Louis, David Trejo, and Catarina Pestana. 2021. "Potential Influencing Factors Related to Digitalization of Construction-Phase Information Management by Project Owners." *Journal of Management in Engineering* 37 (3): 04021010. https://doi.org/10.1061/(ASCE)ME.1943-5479.0000903.

Järvinen, Pertti. 2007. "Action Research Is Similar to Design Science." *Quality & Quantity* 41 (1): 37–54. https://doi.org/10.1007/s11135-005-5427-1.

Jiang, Feng, Ling Ma, Tim Broyd, and Ke Chen. 2021. "Digital Twin and Its Implementations in the Civil Engineering Sector." *Automation in Construction* 130 (October): 103838. https://doi.org/10.1016/j.autcon.2021.103838.

Kritzinger, Werner, Matthias Karner, Georg Traar, Jan Henjes, and Wilfried Sihn. 2018. "Digital Twin in Manufacturing: A Categorical Literature Review and Classification." *IFAC-PapersOnLine*, 16th IFAC Symposium on Information Control Problems in Manufacturing INCOM 2018, 51 (11): 1016–22. https://doi.org/10.1016/j.ifacol.2018.08.474.

Lin, Yu-Cheng, and Weng-Fong Cheung. 2020. "Developing WSN/BIM-Based Environmental Monitoring Management System for Parking Garages in Smart Cities." *Journal of Management in Engineering* 36 (3): 04020012. https://doi.org/10.1061/(ASCE)ME.1943-5479.0000760.

Love, Peter E. D, Zahir Irani, and David J Edwards. 2004. "Industry-Centric Benchmarking of Information Technology Benefits, Costs and Risks for Small-to-Medium Sized Enterprises in Construction." *Automation in Construction* 13 (4): 507–24. https://doi.org/10.1016/j.autcon.2004.02.002.

Love, Peter E. D., Zahir Irani, and David J. Edwards. 2005. "Researching the Investment of Information Technology in Construction: An Examination of Evaluation Practices." *Automation in Construction*, 20th International Symposium on Automation and Robotics in Construction: The Future Site, 14 (4): 569–82. https://doi.org/10.1016/j.autcon.2004.12.005.

Love, Peter E. D., and Jane Matthews. 2019. "The 'How' of Benefits Management for Digital Technology: From Engineering to Asset Management." *Automation in Construction* 107 (November): 102930. https://doi.org/10.1016/j.autcon.2019.102930.

Love, Peter E. D., Jane Matthews, and Jingyang Zhou. 2020. "Is It Just Too Good to Be True? Unearthing the Benefits of Disruptive Technology." *International Journal of Information Management* 52 (June): 102096. https://doi.org/10.1016/j.ijinfomgt.2020.102096.

Lu, Qiuchen, Ajith Kumar Parlikad, Philip Woodall, Gishan Don Ranasinghe, Xiang Xie, Zhenglin Liang, Eirini Konstantinou, James Heaton, and Jennifer Schooling. 2020. "Developing a Digital Twin at Building and City Levels: Case Study of West Cambridge Campus." *Journal of Management in Engineering* 36 (3): 05020004. https://doi.org/10.1061/(ASCE)ME.1943-5479.0000763.

Lu, Yujie, Yongkui Li, Miroslaw Skibniewski, Zhilei Wu, Runshi Wang, and Yun Le. 2015. "Information and Communication Technology Applications in Architecture, Engineering, and Construction Organizations: A 15-Year Review." *Journal of Management in Engineering* 31 (1): A4014010. https://doi.org/10.1061/(ASCE)ME.1943-5479.0000319.

Madni, Azad M., Carla C. Madni, and Scott D. Lucero. 2019. "Leveraging Digital Twin Technology in Model-Based Systems Engineering." *Systems* 7 (1): 7. https://doi.org/10.3390/systems7010007.

Martinelli, Irene, Federico Campi, Emanuele Checcacci, Giulio Marcello Lo Presti, Francesco Pescatori, Antonio Pumo, and Michele Germani. 2019. "Cost Estimation Method for Gas Turbine in Conceptual Design Phase." *Procedia CIRP*, 29th CIRP Design Conference 2019, 08-10 May 2019, Póvoa de Varzim, Portgal, 84 (January): 650–55. https://doi.org/10.1016/j.procir.2019.04.311.

Milis, Koen, and Roger Mercken. 2004. "The Use of the Balanced Scorecard for the Evaluation of Information and Communication Technology Projects." *International Journal of Project Management* 22 (2): 87–97. https://doi.org/10.1016/S0263-7863(03)00060-7.

Myers, Sumner, Donald George Marquis, and National Science Foundation (U.S.). 1969. *Successful Industrial Innovations: A Study of Factors Underlying Innovation in Selected Firms*. National Science Foundation.

Nam, C. H., and C. B. Tatum. 1992. "Strategies for Technology Push: Lessons from Construction Innovations." *Journal of Construction Engineering and Management* 118 (3): 507–24. https://doi.org/10.1061/(ASCE)0733-9364(1992)118:3(507).

Nemet, Gregory F. 2009. "Demand-Pull, Technology-Push, and Government-Led Incentives for Non-Incremental Technical Change." *Research Policy* 38 (5): 700–709. https://doi.org/10.1016/j.respol.2009.01.004.

Neto, Anis Assad, Fernando Deschamps, Elias Ribeiro da Silva, and Edson Pinheiro de Lima. 2020. "Digital Twins in Manufacturing: An Assessment of Drivers, Enablers and Barriers to Implementation." *Procedia CIRP*, 53rd CIRP Conference on Manufacturing Systems 2020, 93 (January): 210–15. https://doi.org/10.1016/j.procir.2020.04.131.

Nguyen, Truong, Li Zhou, Virginia Spiegler, Petros Ieromonachou, and Yong Lin. 2018. "Big Data Analytics in Supply Chain Management: A State-of-the-Art Literature Review." *Computers & Operations Research* 98 (October): 254–64. https://doi.org/10.1016/j.cor.2017.07.004.

Opoku, De-Graft Joe, Srinath Perera, Robert Osei-Kyei, and Maria Rashidi. 2021. "Digital Twin Application in the Construction Industry: A Literature Review." *Journal of Building Engineering* 40 (August): 102726. https://doi.org/10.1016/j.jobe.2021.102726.

Oyegoke, Adekunle Sabitu, and Juhani Kiiras. 2009. "Development and Application of the Specialist Task Organization Procurement Approach." *Journal of Management in Engineering* 25 (3): 131–42. https://doi.org/10.1061/(ASCE)0742-597X(2009)25:3(131).

Peffers, Ken, Tuure Tuunanen, Marcus A. Rothenberger, and Samir Chatterjee. 2007. "A Design Science Research Methodology for Information Systems Research." *Journal of Management Information Systems* 24 (3): 45–77. https://doi.org/10.2753/MIS0742-1222240302.

Peppard, Joe. 2016. "What about the Benefits? A Missing Perspective in Software Engineering." In *Proceedings of the 10th ACM/IEEE International Symposium on Empirical Software Engineering and Measurement*, 1. ESEM '16. New York, NY, USA: Association for Computing Machinery. https://doi.org/10.1145/2961111.2962642.

Perno, Matteo, Lars Hvam, and Anders Haug. 2022. "Implementation of Digital Twins in the Process Industry: A Systematic Literature Review of Enablers and Barriers." *Computers in Industry* 134 (January): 103558. https://doi.org/10.1016/j.compind.2021.103558.

Pyne, Saumyadipta, B.L.S. Prakasa Rao, and S.B. Rao, eds. 2016. *Big Data Analytics*. New Delhi: Springer India. https://doi.org/10.1007/978-81-322-3628-3.

Renkema, Theo J. W. 2000. *The IT Value Quest: How to Capture the Business Value of IT-Based Infrastructure*. https://www.wiley.com/en-us/The+IT+Value+Quest%3A+How+to+Capture+the+Business+Value+of+IT+Based+Infrastructure-p-9780470860557.

Rosen, Roland, Georg von Wichert, George Lo, and Kurt D. Bettenhausen. 2015. "About The Importance of Autonomy and Digital Twins for the Future of Manufacturing." *IFAC-PapersOnLine* 48 (3): 567–72. https://doi.org/10.1016/j.ifacol.2015.06.141.

Rosenberg, Nathan, and Rosenberg Nathan. 1982. *Inside the Black Box: Technology and Economics*. Cambridge University Press.

Schleich, Benjamin, Nabil Anwer, Luc Mathieu, and Sandro Wartzack. 2017. "Shaping the Digital Twin for Design and Production Engineering." *CIRP Annals* 66 (1): 141–44. https://doi.org/10.1016/j.cirp.2017.04.040.

Schmookler, Jacob. 2013. *Invention and Economic Growth*. *Invention and Economic Growth*. Harvard University Press. https://doi.org/10.4159/harvard.9780674432833.

Schroeder, Greyce N., Charles Steinmetz, Carlos E. Pereira, and Danubia B. Espindola. 2016. "Digital Twin Data Modeling with AutomationML and a Communication Methodology for Data Exchange." *IFAC-PapersOnLine*, 4th IFAC Symposium on Telematics Applications TA 2016, 49 (30): 12–17. https://doi.org/10.1016/j.ifacol.2016.11.115.

Shao, Guodong, and Moneer Helu. 2020. "Framework for a Digital Twin in Manufacturing: Scope and Requirements." *Manufacturing Letters* 24 (April): 105–7. https://doi.org/10.1016/j.mfglet.2020.04.004.

Shapiro, Stuart C. 1992. *Encyclopedia of Artificial Intelligence*. 2nd ed. USA: John Wiley & Sons, Inc.

Stockdale, Rosemary, Craig Standing, and Peter E. D. Love. 2006. "Propagation of a Parsimonious Framework for Evaluating Information Systems in Construction." *Automation in Construction*, Knowledge Enabled Information System Applications in Construction, 15 (6): 729–36. https://doi.org/10.1016/j.autcon.2005.09.005.

Succar, Bilal, and Erik Poirier. 2020. "Lifecycle Information Transformation and Exchange for Delivering and Managing Digital and Physical Assets." *Automation in Construction* 112 (April): 103090. https://doi.org/10.1016/j.autcon.2020.103090.

Susto, Gian Antonio, Andrea Schirru, Simone Pampuri, Seán McLoone, and Alessandro Beghi. 2015. "Machine Learning for Predictive Maintenance: A Multiple Classifier Approach." *IEEE Transactions on Industrial Informatics* 11 (3): 812–20. https://doi.org/10.1109/TII.2014.2349359.

Tezel, Algan, Pedro Febrero, Eleni Papadonikolaki, and Ibrahim Yitmen. 2021. "Insights into Blockchain Implementation in Construction: Models for Supply Chain Management." *Journal of Management in Engineering* 37 (4): (ASCE)ME.1943-5479.0000939, 04021038. https://doi.org/10.1061/(ASCE)ME.1943-5479.0000939.

Uhlemann, Thomas H. -J., Christian Lehmann, and Rolf Steinhilper. 2017. "The Digital Twin: Realizing the Cyber-Physical Production System for Industry 4.0." *Procedia CIRP*, The 24th CIRP Conference on Life Cycle Engineering, 61 (January): 335–40. https://doi.org/10.1016/j.procir.2016.11.152.

Van Aken, Joan Ernst. 2005. "Management Research as a Design Science: Articulating the Research Products of Mode 2 Knowledge Production in Management." *British Journal of Management* 16 (1): 19–36. https://doi.org/10.1111/j.1467-8551.2005.00437.x.

Van Der Heijden, Cornelius, and Colin Eden. 1998. "The Theory and Praxis of Reflective Learning in Strategy Making." In , edited by C. Eden and J. C. Spender, 58–76. London: Sage. https://strathprints.strath.ac.uk/43723/.

Varian, Hal R. 2010. "Computer Mediated Transactions." *American Economic Review* 100 (2): 1–10. https://doi.org/10.1257/aer.100.2.1.

Venkatraman, N. 1994. "IT-Enabled Business Transformation: From Automation to Business Scope Redefinition." MIT Sloan Management Review. 1994. https://sloanreview.mit.edu/article/itenabled-business-transformation-from-automation-to-business-scope-redefinition/.

Wache, Hendrik, and Barbara Dinter. 2020. "The Digital Twin – Birth of an Integrated System in the Digital Age." In . https://doi.org/10.24251/HICSS.2020.671.

Walker, Derek H. T., Lynda Margaret Bourne, and Arthur Shelley. 2008. "Influence, Stakeholder Mapping and Visualization." *Construction Management and Economics* 26 (6): 645–58. https://doi.org/10.1080/01446190701882390.

Wishnow, David, Hossein Rokhsari Azar, and Maziar Pashaei Rad. 2019. "A Deep Dive into Disruptive Technologies in the Oil and Gas Industry." In . OnePetro. https://doi.org/10.4043/29779-MS.

Wright, Louise, and Stuart Davidson. 2020. "How to Tell the Difference between a Model and a Digital Twin." *Advanced Modeling and Simulation in Engineering Sciences* 7 (1): 13. https://doi.org/10.1186/s40323-020-00147-4.

Zhou, Cheng, Hanbin Luo, Weili Fang, Ran Wei, and Lieyun Ding. 2019. "Cyber-Physical-System-Based Safety Monitoring for Blind Hoisting with the Internet of Things: A Case Study." *Automation in Construction* 97 (January): 138–50. https://doi.org/10.1016/j.autcon.2018.10.017.